\documentclass[sigconf]{acmart}
\usepackage{subcaption}

\usepackage{listings}
\usepackage{lipsum}
\usepackage{courier}
\usepackage{afterpage}
\newcommand{\ignore}[1]{}
\newif\ifappendix
\appendixtrue
\renewcommand\footnotetextcopyrightpermission[1]{} 

\copyrightinfo{\scriptsize 
\copyright~2020 IBM Corporation}

\newcommand{\numsteps}{seven}

\begin{document}

\title{A Methodology for Creating AI FactSheets}

\author{John Richards, David Piorkowski, Michael Hind, Stephanie Houde, Aleksandra Mojsilovi\'c}


\affiliation{IBM Research}
\begin{abstract}
  As AI models and services are used in a growing number of high-stakes areas, a consensus is forming around the need for a clearer record of how these models and services are developed to increase trust.
  Several proposals for higher quality and more consistent AI documentation have emerged to address ethical and legal concerns
  and general social impacts of such systems.
  However, there is little published work on how to create this documentation.
  This is the first work to describe
  a methodology for creating the form of AI documentation we call FactSheets. We have used this methodology
  to create useful FactSheets for nearly two dozen models.
  This paper describes this methodology and shares the insights we have gathered.   
  Within each step of the methodology, we describe the issues
  to consider and the questions to explore with the relevant people in an organization who will be creating and consuming the AI facts in a FactSheet.
  This methodology will accelerate the broader adoption of transparent AI documentation.  
\end{abstract}

\maketitle

\pagestyle{plain} 

\section{Introduction}
\label{sec-intro}

Recent work has outlined the need for increased transparency in AI for data sets \cite{gebru-2018,data-statements,HollandHNJC2018}, models \cite{model-cards}, and services~\cite{factsheets-2019}. Proposals in support of ethical and trusted AI are also emerging ~\cite{EuropeanCommission2020,raji2019ml,ieee-2017}. Although the specifics differ, all are motivated by the desire to define a set of attributes that capture essential details of how an AI model or service was developed and tested to better understand ethical and legal concerns of the AI system.
Despite the volume of work on transparent reporting mechanisms, there is little work on how to create this documentation. 
Determining \emph{what information} to include and \emph{how to collect} that information is not a simple task.
The lack of methodology for providing this information has hindered adoption of AI documentation in enterprises and regulatory bodies. 
To our knowledge this is the first work that describes a methodology for creating this documentation, which we feel will accelerate its broader adoption.  

Our mechanism for transparent AI documentation, called FactSheets~\cite{factsheets-2019}, takes a more general approach to AI transparency than
previous 
work~\cite{gebru-2018,data-statements,HollandHNJC2018,model-cards,EuropeanCommission2020}
in several ways.
\begin{itemize}
    \item FactSheets are
tailored to the particular AI model or service being documented, and thus can vary in content, 

\item FactSheets are tailored to the needs of their target audience or consumer, and thus can vary in content and format, even for the same model or service,

\item FactSheets capture model or service facts from the entire AI lifecycle, 

\item FactSheets are compiled with inputs from multiple roles in this lifecycle as they perform their actions to increase the accuracy of these facts.
\end{itemize}

We focus on the ability to document the final AI service in additional to an individual model for 3 reasons~\cite{factsheets-2019}.

\begin{itemize}
    \item AI services are the building blocks for many AI applications.
Developers call the service API and
consume its output. An AI service can be an amalgam of many models
trained on many datasets.  Thus, the models and datasets are (direct
and indirect) components of an AI service, but they are not the
interface to the developer.

\item An expertise gap often exists between the producer and consumer of
an AI service. The production team leverages the
creation of one or more AI models and thus will mostly contain data
scientists. The consumers of the API services tend to be developers.
When such an expertise gap exists, it becomes more crucial to communicate the attributes of the artifact in a consumable way.

\item Systems composed of trusted models may not necessarily be trusted, so
it is prudent to also consider transparency and accountability of
services in addition to datasets and models. In doing so, we take a
functional perspective on the overall service and can test for
performance, safety, and security aspects that are not relevant for a
dataset in isolation, such as generalized accuracy, explainability, and adversarial robustness.
\end{itemize}

Our methodology is motivated by user-centered design principles~\cite{user-centered_design}, where user input from multiple stakeholders is collected to inform design. Although this takes more time than a single person designing the documentation, it is significantly more likely to meet the needs of FactSheet consumers.
This paper focuses on a specific form of AI documentation, {\em FactSheets}, however, the techniques will apply to other forms of AI documentation. 

Before we describe our methodology, we first describe a few key concepts.  
Section~\ref{sec-lifecycle} describes the AI lifecycle, summarizing the relevant roles and workflow for the construction and deployment of an AI model or service.
Section~\ref{sec-FactSheets} describes the concept of a FactSheet and motivates the need for a FactSheet Template. Section~\ref{sec-methodology} presents our \numsteps-step methodology for constructing useful FactSheets.
Section~\ref{sec-final} presents further guidance for those organizations planning to create FactSheets.
Section~\ref{sec-practice} provides a concrete example 
FactSheet Template for a model catalog scenario and external user persona.
Section~\ref{sec-harm-safety} concludes by discussing how the methodology can help to improve the needs of consumers with regards to the potential safety and harm of AI.

\section{The AI Lifecycle}
\label{sec-lifecycle}
\begin{figure}
    \centering
    \includegraphics[width=0.45\textwidth]{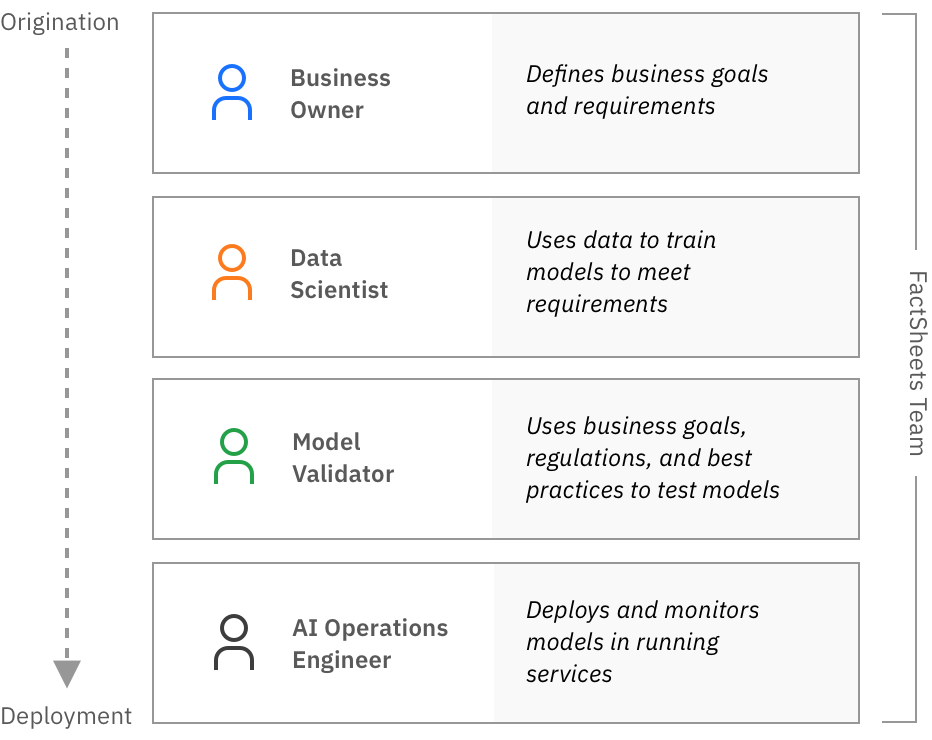}
    \caption{Key roles in a typical AI lifecycle}
    \label{fig:lifecycle}
\end{figure}

The AI lifecycle includes a variety of roles, performed by people with different specialized skills and knowledge that collectively produce an AI service. Each role contributes in a unique way, using different tools. Figure~\ref{fig:lifecycle} specifies some common roles.

The canonical process starts with a business owner who requests the construction of an AI model or service. The request includes the purpose of the model or service, how to measure its effectiveness, and any other constraints, such as bias thresholds, appropriate datasets, or the required levels of explainability and robustness.

The data scientist uses this information to construct a candidate model by using, most typically, a machine learning process. This iterative process includes selecting and transforming the dataset, discovering the best machine learning algorithm, tuning algorithm parameters, etc. The goal is to produce a model that best satisfies the requirements set by the business owner.

Before this model is deployed it often must be tested by an independent person, referred to as a model validator in Figure 1.
This role, often falling within the scope of model risk management~\cite{model-risk-mgmt}, 
third party testing~\cite{3rd-party-testing,eu-testing} or certification~\cite{certification,eu-testing}, is similar to a testing role in traditional software development. 
A person in this role may apply a different test dataset to the model and independently measure metrics defined by the business owner. If the validator approves the model, it can be deployed.

The AI operations engineer is responsible for deploying and monitoring the model in production to ensure it operates as expected. This can include monitoring its performance metrics, as defined by the business owner. If some metrics are not meeting expectations, the operations engineer is responsible for taking actions and informing the appropriate roles.

AI lifecycles will include iteration within a role (a data scientist, building many models before passing it to a validator) or between roles (an operations engineer sending a model back to a data scientist because it is performing poorly).
More sophisticated lifecycles will likely have additional roles. A common pattern is for a model to be combined with other models or human-written code to form a service. In such a case the validator's role may be extended to also validate the full service.

A model is not a static object in the lifecycle, and thus, a FactSheet must incorporate the facts and lineage from all phases of the "life of the model".  This will introduce transparency not only into how the model was built and what it does, but also how it was tested, deployed, and used.

\section{FactSheets and Templates} \label{sec-FactSheets}

\textbf{FactSheets}~\cite{factsheets-2019} are a collection of information 
about how an AI model or service was developed and deployed.  
FactSheets summarize the key characteristics of a model or service for use by a variety of stakeholders. We have previously summarized the difficulties developers face when creating FactSheets~\cite{experiences-2020}. 
This paper describes the best practices we have developed in the process of creating FactSheets for nearly two dozen models.
These include FactSheets for standalone models as well as services that encapsulate one or more models.
They cover a wide range of application areas including text analysis and generation, language translation, object detection, object classification in two-dimensional images, audio signal classification, weather forecasting, agricultural crop yield prediction, and facility energy optimization.   

This work has demonstrated that although FactSheets will contain some common elements, different FactSheets will generally contain different information, at different levels of specificity, depending on domain and model type. They will also contain different information for different industries and the different regulatory schemes within which these industries operate. 

Within a particular domain or organization, FactSheets will also take on different forms, and contain different content, for different purposes. Model validators may need detailed information on data selection and cleaning, feature engineering, and accuracy and bias metrics. Business owners may need information on whether a deployed model is meeting business needs. Regulators may need a report detailing how a model complies with established practices and metrics related to safety, bias, and harm.
Thus, although there is a strong desire to create a standard template for all FactSheets, we believe this diversity illustrates that for FactSheets, \textbf{one size does not fit all}.

We believe that standards will eventually emerge and, like nutrition labels, be useful for some purposes. 
In the foreseeable future, however, many kinds of FactSheets will be created. 
We have created the notion of \textbf{FactSheet Templates} to manage this diversity. A FactSheet Template can be thought of as specifying the categories or types of information that will be collected and displayed during, and and even after, AI development. Any given lifecycle will likely have multiple templates since different people will likely want to see different information, for different purposes, at different points in time. A large part of the job of creating FactSheets is designing the appropriate FactSheet Template(s). This will be a prime focus of Section~\ref{sec-methodology}.

\section{FactSheet Methodology}
\label{sec-methodology}
\begin{figure}
    \centering
    \includegraphics[width=0.45\textwidth]{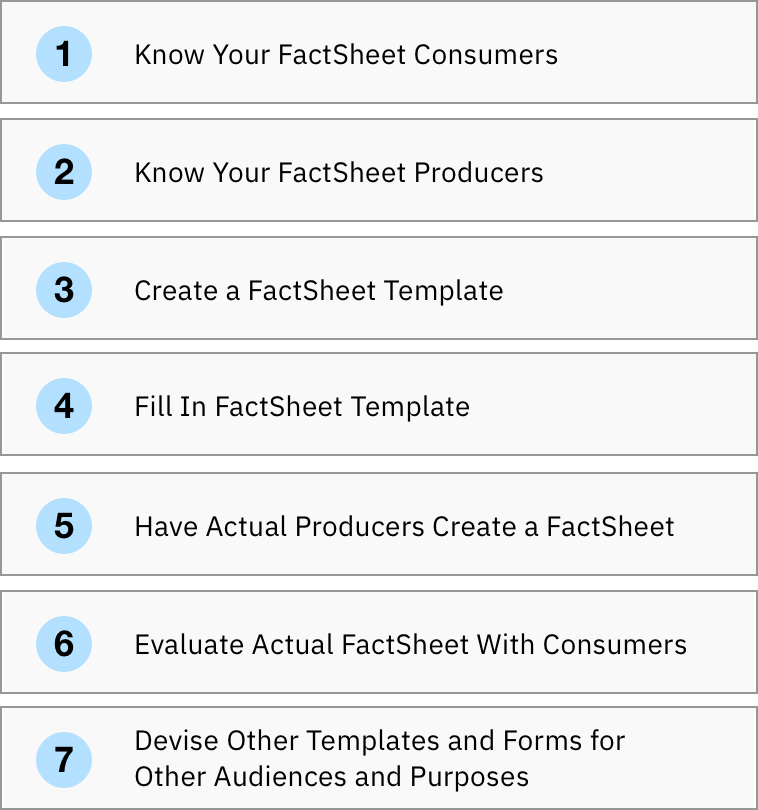}
    \caption{Steps to produce useful FactSheets}
    \label{fig:flowchart}
\end{figure}

We now describe our \numsteps-step methodology for the construction of useful FactSheets.
For expository purposes, the steps shown in Figure~\ref{fig:flowchart} are presented as though they flow in an uninterrupted stream from beginning to end. The reality is that FactSheet production is highly iterative, especially in the early days of FactSheet adoption within an organization.

Each step lists the key roles involved. In addition to the more typical roles shown in Figure~\ref{fig:lifecycle}, an additional role is identified, namely the ``FactSheets Team". This team is responsible for designing and implementing the FactSheets process within the organization. The first three steps will be driven by this team as they interview potential FactSheet consumers and producers and design the first FactSheet Template. Step 4 will largely be performed by the FactSheets Team but will benefit from the involvement of those with direct knowledge of the model or service being documented. This step may involve several iterations and informal trials with potential consumers and producers. In Step 5, FactSheet producers will be generating an actual FactSheet. In Step 6, FactSheet consumers will be assessing the quality and usefulness of this FactSheet. The FactSheets Team will be involved in these latter steps as well but will rely heavily on others to produce and attempt to consume actual content. In Step 7, the FactSheets Team repeats the process to increase coverage and value.

To simplify the presentation in the following steps we focus on one fact producer, ``Priya", and one fact consumer ``Carmen". Priya is a data scientist who will generate facts about how she created her model. Carmen is a model validator who will assess the model Priya created on various dimensions including quality, simplicity, and potential risk. Of course, Priya may also be a consumer of facts produced earlier by those who assembled the training data she uses. Similarly, Carmen may be a producer of facts for those who make the final decision on deployment readiness of the model she validates.

This may seem like a lot to think about, especially when there are multiple roles to understand and a desire to broadly sample multiple representative users within each role. But the important thing is to start. Find one person performing each role (some people will be performing more than one role). Spend 30 minutes in conversation with each of them. If needed, find more than one person to explore areas that are still unclear after the first conversation. To speed things up, consider bringing potential producers and consumers together in conversation at any point in this process. They may quickly converge on what information is needed and how it can be produced in a cost-effective way. 

\subsection{Step 1: Know Your FactSheet Consumers}

\begin{itemize}
    \item Who: FactSheets Team (with potential consumers)
    \item What: Gather the information needs of potential FactSheet consumers
\end{itemize}

FactSheets are produced so that they can be consumed. Understanding the information needs of FactSheet consumers is the first and most important task. Here are some of the questions to consider in this first step (with Carmen, a model validator, as the illustrative consumer):

\begin{enumerate}
\item What does Carmen do now when she performs her role?
\item What is Carmen going to be asking for when looking at a FactSheet?
\item What decisions will she be making based on the information presented?
\item How is the FactSheet going to help her do her job more effectively?
\item What are the most important pieces of information that Carmen needs to know?
\item What is Carmen's level of expertise in general data science?
\item How is Carmen's expertise going to affect the information presented?
\item Will there need to be additional definitions for terms that Carmen is unfamiliar with?
\item What is Carmen's level of expertise with respect to the model algorithms being used?
\item What explanations about the model's algorithm or results is Carmen going to need?
\item What is Carmen's level of expertise in the problem domain?
\item How is that going to affect the information presented?
\item Will Carmen need help in mapping general knowledge of the problem domain to the particular inputs, outputs, or performance indicators associated with this model?
\item Is Carmen aware of issues related to model risk, potential harm, and regulatory compliance?
\item What information is needed assess these issues?
\end{enumerate}

\subsection{Step 2: Know Your FactSheet Producers}

\begin{itemize}
    \item Who: FactSheets Team (with potential producers)
    \item What: Gather the kinds of information FactSheet producers might generate
\end{itemize}

Some facts can be automatically generated by tooling. Some facts can only be produced by a knowledgeable human. Both kinds of facts will be considered during this step. Here are some of the questions we might explore with Priya (a data scientist) about the facts she could usefully generate during the creation of a model:

\begin{enumerate}
\item What facts does Priya wish she could conveniently record about the models she develops? It is often helpful to ask about the most recent model, or a model that was particularly important, or a model that was exceptionally difficult to produce, rather than discussing models in general.
\item What did Priya do during the creation of this model that is otherwise unknown to others?
\item Are there general facts about the data, the features, the model algorithm, or the training and testing Priya performs that are important to note? Why?
\item What model-specific knowledge does she have that may not be obvious to others?
\item What domain-specific knowledge does Priya have that may not be obvious to others?
\item Does Priya know who will be consuming the facts she produces? We will assume it is Carmen in this particular case. Does Priya know Carmen? Have they talked about what Carmen needs to know?
\item Is Priya aware of issues related to model risk, potential harm, and regulatory compliance?
\item What information will be needed by others to assess these issues?
\end{enumerate}

\subsection{Step 3: Create a FactSheet Template}

\begin{itemize}
    \item Who: FactSheets Team
    \item What: Define the topics and questions to be included in FactSheets
\end{itemize}

What is learned in these first two steps leads directly to the most important part of creating FactSheets, namely the creation of a FactSheet Template. As discussed in Section~\ref{sec-FactSheets}, a FactSheet Template will contain what can be thought of as questions. Each individual FactSheet will contain the answers to these questions. For example a template may start with the question "What is this model for?". It may then expand on that by asking about where the model is well suited and where the model is ill suited.

The information gathered in the first two steps will inform the creation of this FactSheet Template. You may find that details about how a model is created are much less important in your organization than information about risk assessments and regulatory compliance. Or you may find that detailed questions about robustness against adversarial attacks is needed because of the nature of the models you create or the high-stakes domains within which they are used.

Here are some of the questions to consider in creating the first iteration of a FactSheet Template. Again, this is cast in terms of Carmen's needs for information and Priya's ability to produce that information, but similar questions will apply to many of the roles in the AI lifecycle
or external consumers of the AI documentation.

\begin{enumerate}
    \item What are the topics or categories of information needed?
    \item Do some of these categories have subcategories?
    \item What is a meaningful name for each category or subcategory?
    \item What kinds of information should be included in each category? For example, Carmen may want to group all the model performance metrics within a category called "Model Performance". Information about the representativeness of the training data might be grouped with information on the sensitivity of the model to drift in a category called, 'Potential Sources of Error'.
    \item How should each question in a category be worded so as to be both understandable and evocative for Priya? The goal here is to encourage fact producers to answer in ways that are concise, germane, and understandable.
    \item Where will the answer to a question come from? Will it be generated automatically by a tool or entered by a knowledgeable human? If the former, will Priya have some control over the frequency of fact generation or the granualarity of recorded facts? If the latter, will Priya be given hints or examples of the kind of answer that would be satisfactory?
    \item Are there any regulatory, legal, or business concerns that need to be considered when answering the questions in this template?
    \item Are there different presentation formats needed for this information (for example, a short tabular summary of just key facts, or a slide format for presentations to review boards)? AI FactSheets 360~\cite{fs360} shows three different formats that might be useful.
    \item In addition to the human-readable content, is there a need for machine-readable content that Priya might generate?
\end{enumerate}

\subsection{Step 4: Fill In FactSheet Template}

\begin{itemize}
    \item Who: FactSheets Team
    \item What: Informally assess FactSheet Template by trying to fill it in
\end{itemize}

This step is where you will attempt to fill in your FactSheet Template for the first time. As you do this, informally assess the quality of the template itself. While this assessment is not a substitute for further work with Priya and Carmen (to follow), it may quickly highlight where improvements are needed. In doing this assessment try to reflect on the template, and the FactSheets it will generate, from Carmen's and Priya's points of view. Ask yourself, or other members of your FactSheets Team, the following questions:

\begin{enumerate}
\item Knowing what Carmen knows, will she be able to understand the information that filled-in FactSheets will include?
\item Are there details needed by Carmen that will be missing in these FactSheets?
\item Is there specialized language that Carmen will be unfamiliar with?
\item Will the information allow Carmen to make the decisions she needs to make?
\item How are these FactSheets going to help Carmen do her job more effectively?
\item What might we do to encourage Priya to answer questions in ways that provide what Carmen needs?
\end{enumerate}

\subsection{Step 5: Have Actual Producers Create a FactSheet}

\begin{itemize}
    \item Who: Business Owner, Data Scientist, Model Validator, AI Operations Engineer (and others as defined within your organization's AI lifecycle)
    \item What: Populate a FactSheet Template with actual facts
\end{itemize}

At this point you have a solid template and a good sense of how it might be used to create FactSheets. The next step is to have actual fact producers fill in the template for their part of the lifecycle. If there is a question in the template about model purpose, find someone who would actually be entering that information and have them answer the question. Ask a data scientist to answer the questions related to the development and testing of an actual model. If this model was validated, ask the model validator to enter information about that process. 
Similarly, have a person responsible for model deployment answer those questions.
If the lifecycle is not that structured, have the person responsible for most of the work create this FactSheet. 

We have found this step to be highly iterative. You can expect sections of your template to be expanded, compressed, or eliminated altogether. Individual questions will be refined within these sections. Stay alert for ideas or helpful hints about other fact producers that may surface. Follow these leads later. The goal here is to create a FactSheet that is ready for evaluation by consumers in the next step. Take the time to get this FactSheet to a level of quality and completeness that will make this next evaluation meaningful.

\subsection{Step 6: Evaluate Actual FactSheet With Consumers}

\begin{itemize}
    \item Who: Business Owner, Data Scientist, Model Validator, AI Operations Engineer (and others as defined within your organization's AI lifecycle)
    \item What: Assess FactSheet quality with those who will be consuming FactSheets in production
\end{itemize}

In this step we conduct an assessment of the quality and completeness of the actual FactSheet produced in the previous step. If the FactSheet is intended to be used by multiple roles (not uncommon), evaluate it separately for each role. To make each evaluation meaningful, ensure you have agreement with respect to the purpose of the FactSheet. Ask the consumer to imagine using this FactSheet to actually perform their work.

Each evaluation consists of two parts. The first focuses on the content in the FactSheet. The second focuses on the way in which information is presented.

Content Evaluation: The goal of this part of the evaluation is to see how well the content of the FactSheet meets the specifically-designed-for information needs of the consumer. Ask your consumer to go through the FactSheet item by item with their information needs in mind and identify the following:

\begin{enumerate}
    \item What information is missing?
    \item Why is that missing information important to include?
    \item How would they like this information presented?
    \item Can they give an example?
    \item What information is extraneous?
    \item Why is that information extraneous?
    \item What information is confusing or hard to understand?
    \item Why is that information hard to understand?
    \item How can that information be made more understandable?
    \item Can they give an example?
    \item Was the organization of information sensible?
    \item If not, what would they change?
\end{enumerate}

Have the consumer rank the information presented in this FactSheet from most important to least important. Remember to include the information that was noted as missing in this ranking. If time permits, have them share their views about the FactSheet with your larger group. Encourage discussion and ask questions about any unexpected findings, which can often identify gaps in the underlying lifecycle process or confusion about roles. Addressing these gaps can pay large dividends.

Presentation Evaluation: The goal of this part of the evaluation is to see if the way that information is \textit{presented} meets the specifically-designed-for information needs of the consumer. Since some of the information you collect may be visual, make sure to allow for that type of feedback. Ask each consumer to go through the FactSheet item by item with their information needs in mind and identify these things:

\begin{enumerate}
    \item Is this information presented in an unexpected way?
    \item How can the information be presented differently?
    \item Why is this alternative a better way to present this information?
    \item Can they draw or describe an example?
    \item If the information presentation includes interactive elements, are they useful?
    \item How can they be made more useful?
    \item Why is that more useful?
    \item If they could add or change the way that information is presented, how would they?
    \item Why is this addition or change an improvement?
    \item Is this, overall, the right format for presenting this information?
    \item What format would be more suitable?
    \item Why is that format more suitable?
\end{enumerate}

\subsection{Step 7: Devise Other Templates and Forms For Other Audiences and Purposes}

\begin{itemize}
    \item Who: FactSheets Team (and others as appropriate)
    \item What: Evolve existing templates and create new ones
\end{itemize}

By now you will have created a refined FactSheet Template for use by others. They will be able to create useful and consumable FactSheets with that template. But there is more to do. There may be other consumers that need to be supported. Perhaps it is time to turn from an inward focus to an outer one, crafting templates for FactSheets to be consumed by external review boards or regulators. Or it may be time to support other stakeholders not directly involved in the AI lifecycle, such as sales personnel or the ultimate consumers of an AI service. Other formats for the same content may need to be created as well. The above steps can be followed once again. You will have  learned a surprising amount about \textit{how} to create FactSheet Templates and FactSheets from having gone through this process once. It will go faster and more smoothly now.

We encourage an ongoing process of reflecting on how well FactSheets support your AI lifecycle once they are fully incorporated and in routine use. Consider how they might be improved. Perhaps a new business opportunity in a new domain has developed or new types of models are being created that capitalize on new algorithmic research. If so, it may be time to refine existing FactSheet Templates or create new ones.

\section{Further Guidance}
\label{sec-final}

We have observed ~\cite{experiences-2020} that producers of FactSheets have a hard time imagining what consumers of FactSheets need to know and how best to provide that information. Model developers, for example, may have a sophisticated understanding of the algorithmic basis for a model, but may describe the model or its performance in ways that assume far too much knowledge on the part of a FactSheet consumer. Consumers may not really know what information they need to support their work without somewhat structured reflection. Our methodology addresses these gaps by applying a user-centered design process \cite{user-centered_design} to the task of creating useful AI documentation. This process need not be time consuming and expensive. Even talking with a few potential FactSheet consumers and producers will be helpful.

It may be obvious that following this methodology will not cause FactSheets across the vast array of adopting organizations to converge on a single template or a single format. The methodology \textit{will} lead, however, to FactSheets that fit the needs of a particular organization and provide real value to the corresponding AI development, deployment, and monitoring teams.

As noted above, one size will not fit all, at least if you dive below a short nutrition-label-like form to something that provides useful detail to all the lifecycle roles in a real organization. Even FactSheets developed with the same template will differ in interesting ways. For example, some models will have FactSheets with extensive sections on bias and fairness testing with respect to protected populations. Other models will have FactSheets for which fairness and bias considerations are truly not applicable. Within some regulated industries, FactSheets may run to a hundred or more pages while the FactSheet produced by a startup providing an AI component for text sentiment analysis may be little more than a statement of purpose, inputs, and outputs.

This methodology for creating FactSheets may seem like a lot of work. Following these steps \textit{will} take more time than just having a single person write a FactSheet Template based on a limited understanding of the actions and information needs within your organization. But failing to perform these steps will incur an ongoing price in poor documentation, repeated requests between team members for missing information, insufficient testing based on faulty assumptions about data or model structure, suboptimal business results, and exposure to unnecessary risk.

We have found that following these steps with even a small number of people, where
there is perhaps only one representative for each stage in a lifecycle, will pay dividends. We have also found that
iterating quickly, rather than spending substantial time trying to attain perfection within each iteration, will shorten the overall time needed.

\section{In Practice}
\label{sec-practice}

In this section we describe a FactSheet Template created using this methodology. Table~\ref{tbl:max} shows the relatively compact template we developed to create FactSheets for inclusion in the Model Asset eXchange (MAX) open-source catalog ~\cite{MAX}. The intended user of these FactSheets is a developer, examining models in the catalog for possible adoption as part of a larger service or system. In this scenario, the developer needs to know, first, whether the model is suitable for the kind of service they are tying to build. Answers to questions 1, 2, 8, and 9 provide this information. The developer might then want to understand how the model was trained. If it must be retrained for their purposes, 
they need to understand 
what sorts of data the model might use. 
The answer to question 3 will provide this information. 
The answer to question 5 clarifies what input(s) the model receives and what output(s) it produces. This allows the developer to gauge the amount of work required to plug the model into their application framework. Information on how the model was tested, including details about the test data, and the model's performance metrics, are provided by the answers to questions 6 and 7. If needed, information about the model itself, including pointers to papers with further details, is provided by the answer to question 4. Finally, the answer to question 10 allows them to see if they can provide explanations of why a particular output was generated for a particular input. 

An example of a FactSheet produced for a MAX model with this template is shown in Figures 3--12. After examining this FactSheet, the developer is able to decide whether to use this model or keep searching for another one. They are also able to assess the expected performance of the model both initially (accuracy) and over time (whether it is subject to drift). As this model classifies images, and since models for this domain are vulnerable to adversarial attacks, the detailed robustness section may be particularly relevant. Other examples of FactSheets using this template can be found in ~\cite{fs360}.

\begin{table}[H]
\caption{FactSheet Template for MAX Models}
\label{tbl:max}
\centering
\begin{tabular}{|l|} \hline
\\(1) What is this model for?\\
(2) What domain was it designed for?\\
(3) Can you describe information about the training data\\
\hspace{1em} (if appropriate)?\\
(4) Can you provide information about the model\\
\hspace{1em} (if appropriate)?\\
(5) What are the model's inputs and outputs?\\
(6) What are the model's performance metrics?\\
\hspace{2em}- accuracy\\
\hspace{2em}- bias\\
\hspace{2em}- robustness\\
\hspace{2em}- domain shift\\
(7) Can you provide information about the test set?\\
(8) In what circumstances does the model do particularly\\
\hspace{1em} well (within expected use cases of the model)?\\
\hspace{1em} (e.g., inputs that work well)\\
(9) Based on your experience, in what circumstances \\
\hspace{1em} does the model perform poorly? (e.g. domain shift,\\
\hspace{1em} specific kinds of input, observations from experience)\\
(10) Can a user get an explanation of how your model\\
\hspace{1.5em} makes its decisions?\\
\\
\hline
 \end{tabular}
 \end{table}
 
To carry this scenario forward, imagine the developer has incorporated the Object Detector model described in Figures 3--12 into a service designed to categorize photos of damage to property sent from mobile devices. The service would first detect where objects were in the scene, categorize the nature of the damage (e.g., graffiti, broken glass, open hydrants), and use the GPS location in the mobile device to dispatch the right repair crew to the right location.

Imagine, also, that a municipality has created an AI Ethics Review Board to assess the appropriateness of deploying AI systems such as this one. It is clear that a different FactSheet Template, and different FactSheets, would be needed for this use case. 

Table~\ref{tbl:muni} shows a portion of what this template might contain. Some of the questions might be the same as Table~\ref{tbl:max} but would refer to deployable services rather than individual models. Some topics might still be covered but with much less detail. For example, question 4 might ask only for the class of model (for purposes of mapping to a general risk scale) but no further details on model architecture would be needed. Others will require answers that might have been optional for internal FactSheets because of well-established internal controls. Data privacy would be an obvious concern for this Review Board since the smartphone capturing and transmitting the image could probably be traced back to an individual owner. Information about data handling and personally identifiable information might be detailed in the answer to question 5. If deployed within the European Union, compliance with the General Data Protection Regulation (GDPR) ~\cite{gdpr} would be addressed in question 6. How well this service would perform when confronted with poor lighting, busy compositions, and odd angles in the photos provided by untrained users might also be a concern. The sorts of test data that are typical in the image recognition domain may not be adequate to cover this variation. Answers to question 2, 3, and 7 might provide information needed to assess whether this is a valid concern. Subtle questions of potential bias might also surface. Perhaps there are neighborhoods in which crowded streets make the photos of damage hard for the model to correctly detect and classify. Might this lead to more repairs being scheduled for more affluent areas that experience less crowding? Is this fair? The answers to question 8 might provide insight into these potential problems.

This section illustrates how our methodology was used to create a
FactSheet template for a particular use case (model catalog), how
the template was used to complete a FactSheet for a particular model
(Object Detector), and how the same model being part of a different
use case (AI Ethics Board) would lead to a different FactSheet Template.
This demonstrates how the flexibility of the methodology can lead to
more useful transparent reporting mechanisms.

\begin{table}[H]
\caption{Portion of FactSheet Template for an AI Ethics Review Board}
\label{tbl:muni}
\centering
\begin{tabular}{|l|} \hline
\\(1) What does this service do?\\
(2) Provide details about training data including distributions\\
(3) Provide details about the test data including distributions\\
(4) What classes of model are used in the service?\\
(5) Describe data handling protocols in detail\\
(6) Describe GDPR compliance in detail\\
(7) What kinds of inputs will be handled poorly?\\
(8) Describe all issues of possible bias and fairness (even if\\
\hspace{1em} there are no protected attributes in the training data)\\
(9) ...\\
\\
\hline
 \end{tabular}
 \end{table}

\section{Harm and Safety}
\label{sec-harm-safety}
The increasing use of AI systems in high-stakes decision making has
underscored the importance of transparent reporting mechanisms. These mechanisms, including FactSheets, can lead to better understanding, and more effective mitigation of any harm or safety issues in the system, such as bias, vulnerabilities to adversarial attacks, or other undesirable societal impacts. 
For example, a section that describes a detailed analysis of bias in the training dataset can help illuminate if the system is appropriate for a particular use case.


This paper describes a methodology for producing a useful transparent reporting mechanism for AI systems.  This methodology can contribute to the identification of potential harm and safety issues. The methodology does this by 
\begin{itemize}
\item explicitly including multiple FactSheet consumers and producers in FactSheet requirements gathering (Steps 1--2) 
\item asking questions about their concerns for harm and risk (Steps 1--2) 
\item providing a feedback mechanism to allow further input (Step 6)
\item including a broad range of perspectives in the development of FactSheets (Steps 1--7)

\end{itemize}

This process will increase the likelihood that FactSheets will provide the information needed to understand and mitigate potential harm or safety issues with an AI system.

\section*{Acknowledgements}
We thank Karthik Muthuraman, Saishruthi Swaminathan, and the MAX model team for their collaboration in creating FactSheets for MAX models. We also thank Fernando Carlos Martinez for his efforts in creating the FactSheets 360 website~\cite{fs360}, Kush Varshney for his guidance
on the project and feedback on this paper, Kevin Eykholt, Taesung Lee, and Ian Molloy for their analysis of model adversarial robustness, and Noel Codella, Leonid Karlinsky, Brian Kingsbury, Youssef Mroueh, Inkit Padhi, and John Williams for their expert guidance on FactSheet content.

\newpage
\begin{figure*}
\centering
    \includegraphics[width=0.95\linewidth]{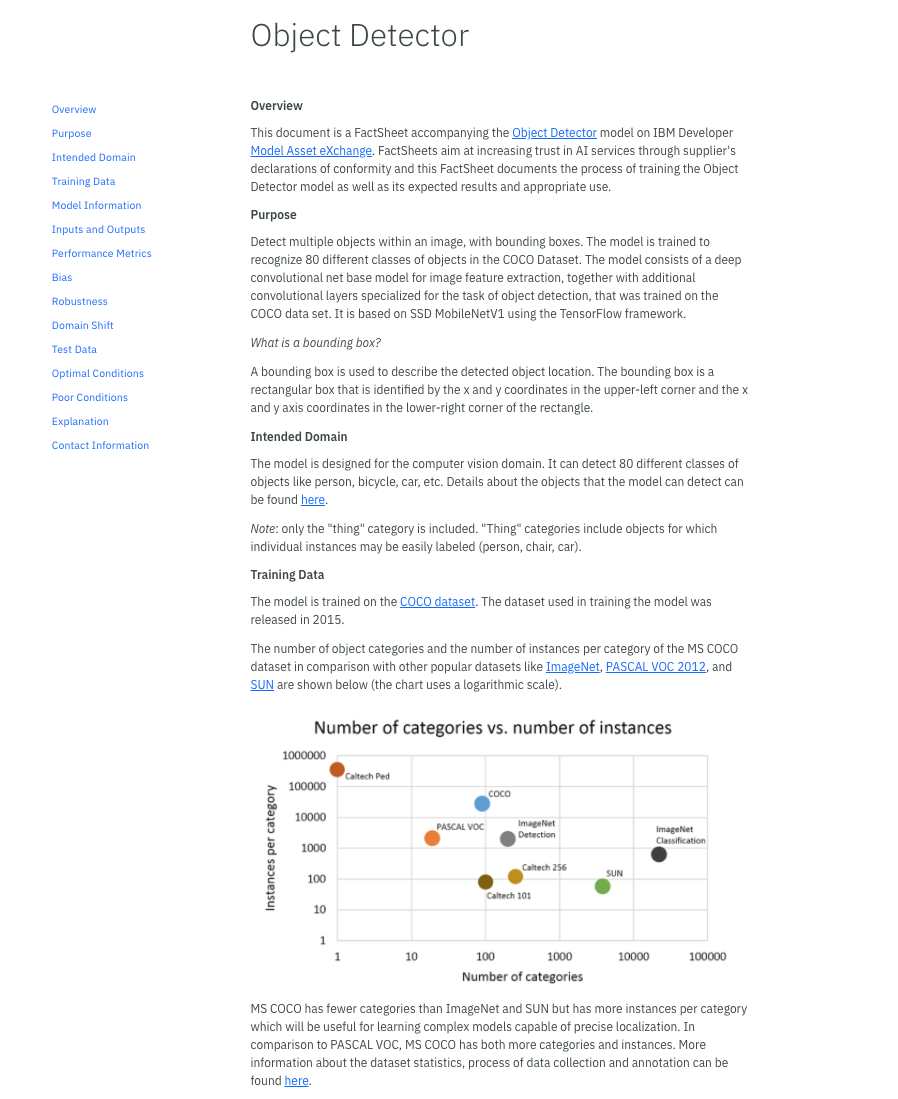}\hfil
\caption{Object Detector FactSheet - 1}
\end{figure*}

\newpage
\begin{figure*}
\centering
    \includegraphics[width=0.84\linewidth]{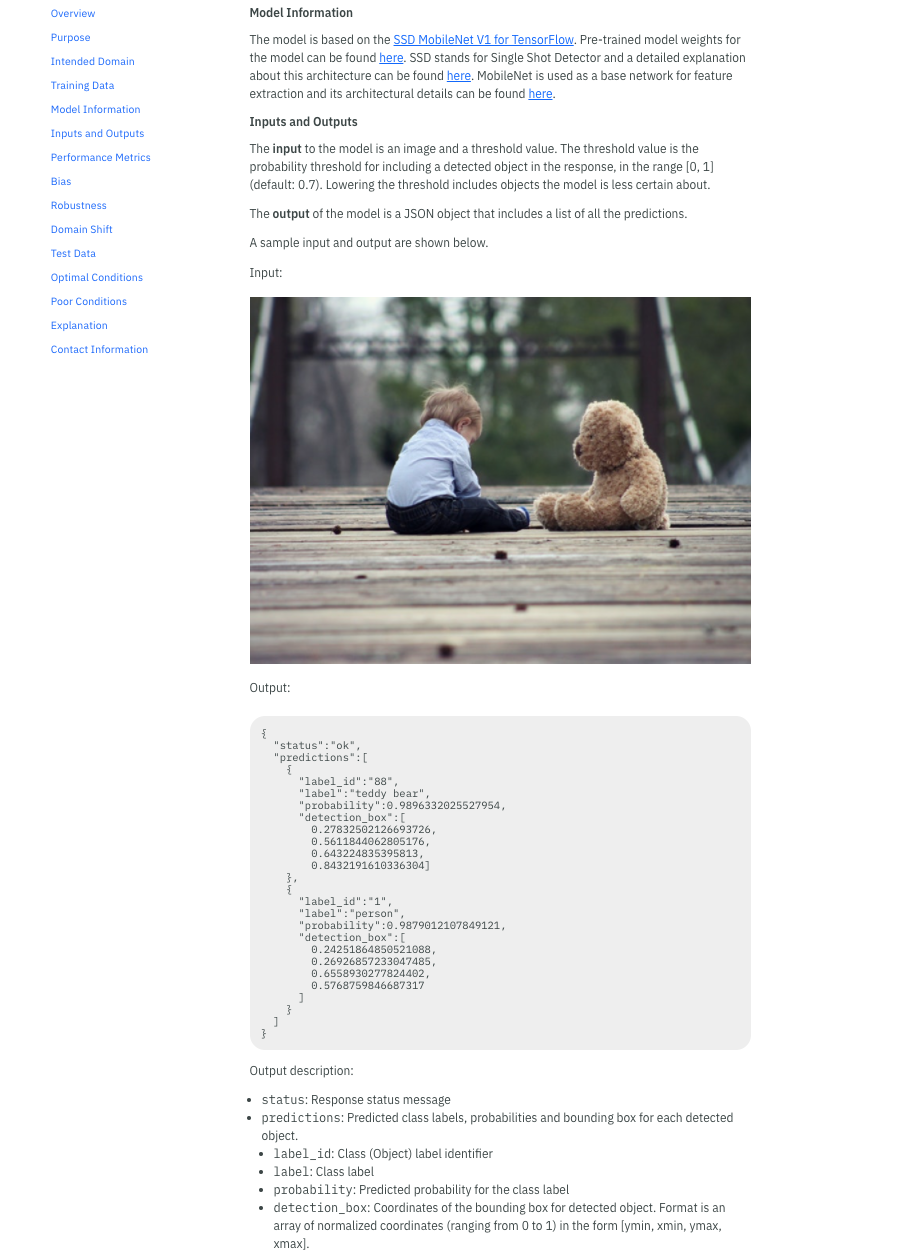}\hfil
\caption{Object Detector FactSheet - 2}
\end{figure*}

\newpage
\begin{figure*}
\centering
    \includegraphics[width=0.95\linewidth]{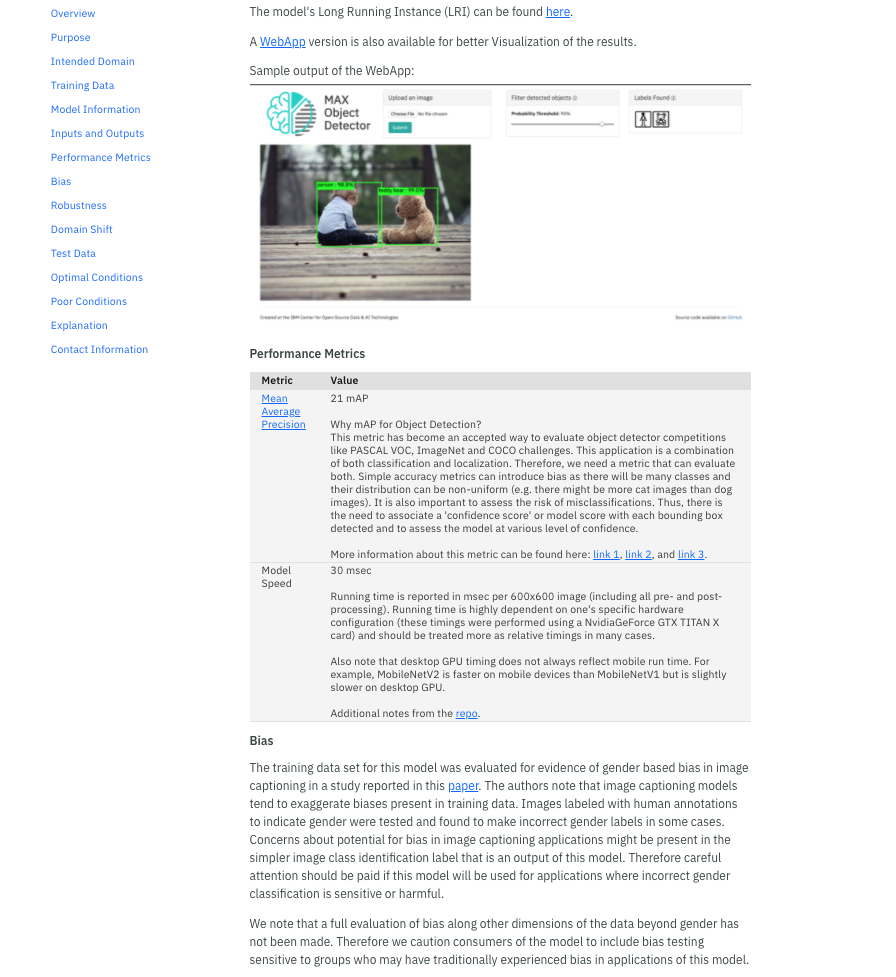}\hfil
\caption{Object Detector FactSheet - 3}
\end{figure*}

\newpage
\begin{figure*}
\centering
    \includegraphics[width=0.95\linewidth]{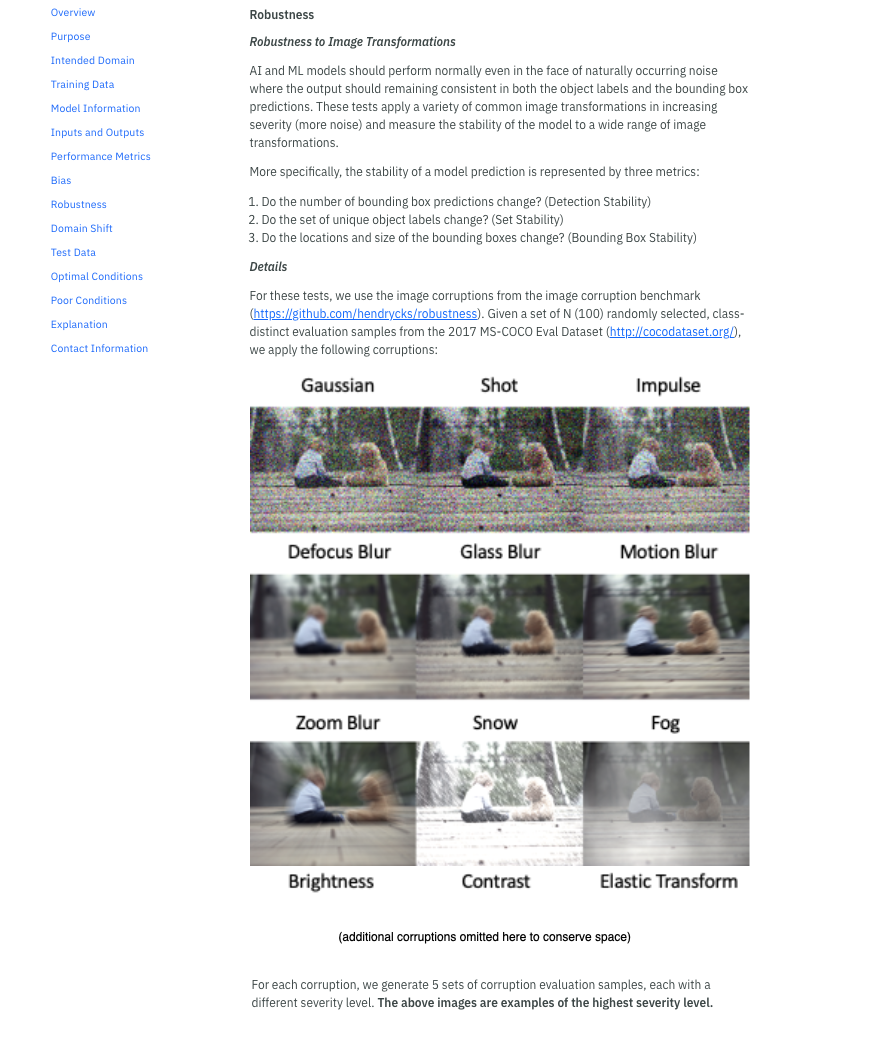}\hfil
\caption{Object Detector FactSheet - 4}
\end{figure*}

\newpage
\begin{figure*}
\centering
    \includegraphics[width=0.95\linewidth]{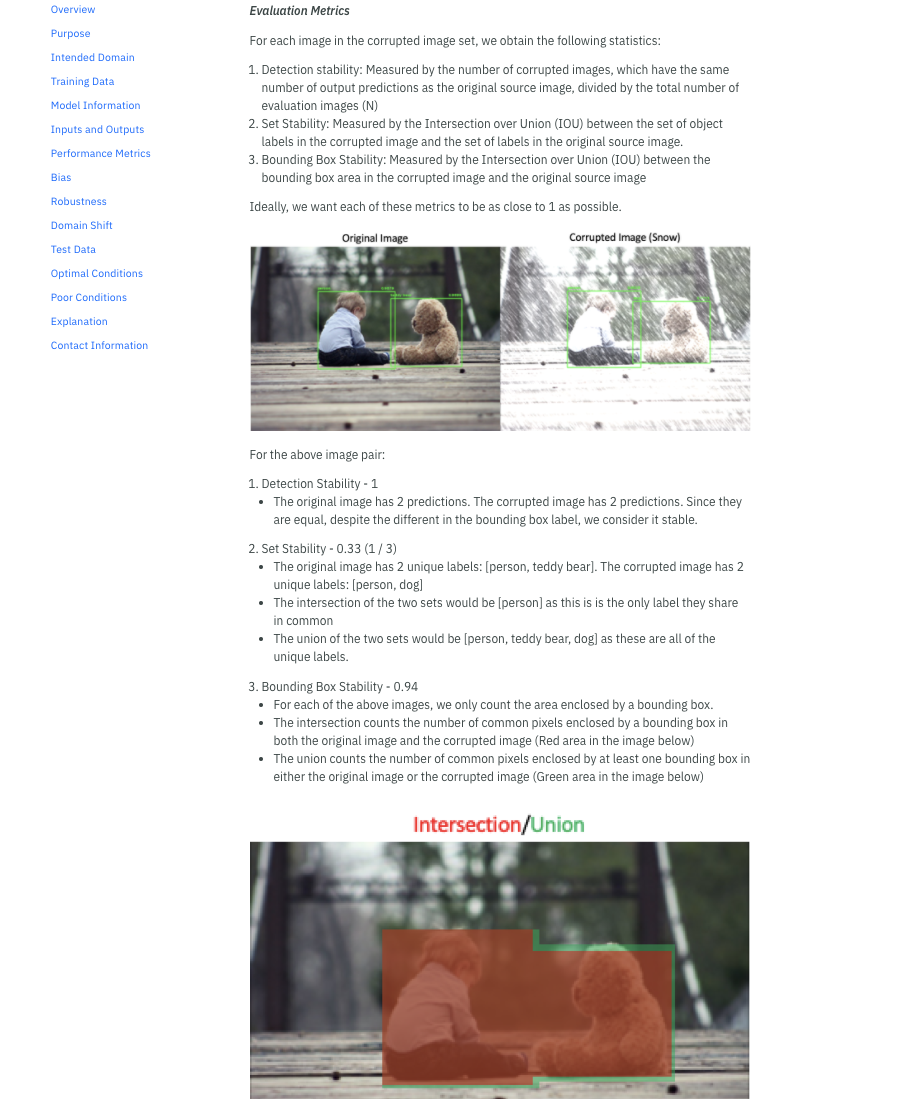}\hfil
\caption{Object Detector FactSheet - 5}
\end{figure*}

\newpage
\begin{figure*}
\centering
    \includegraphics[width=0.95\linewidth]{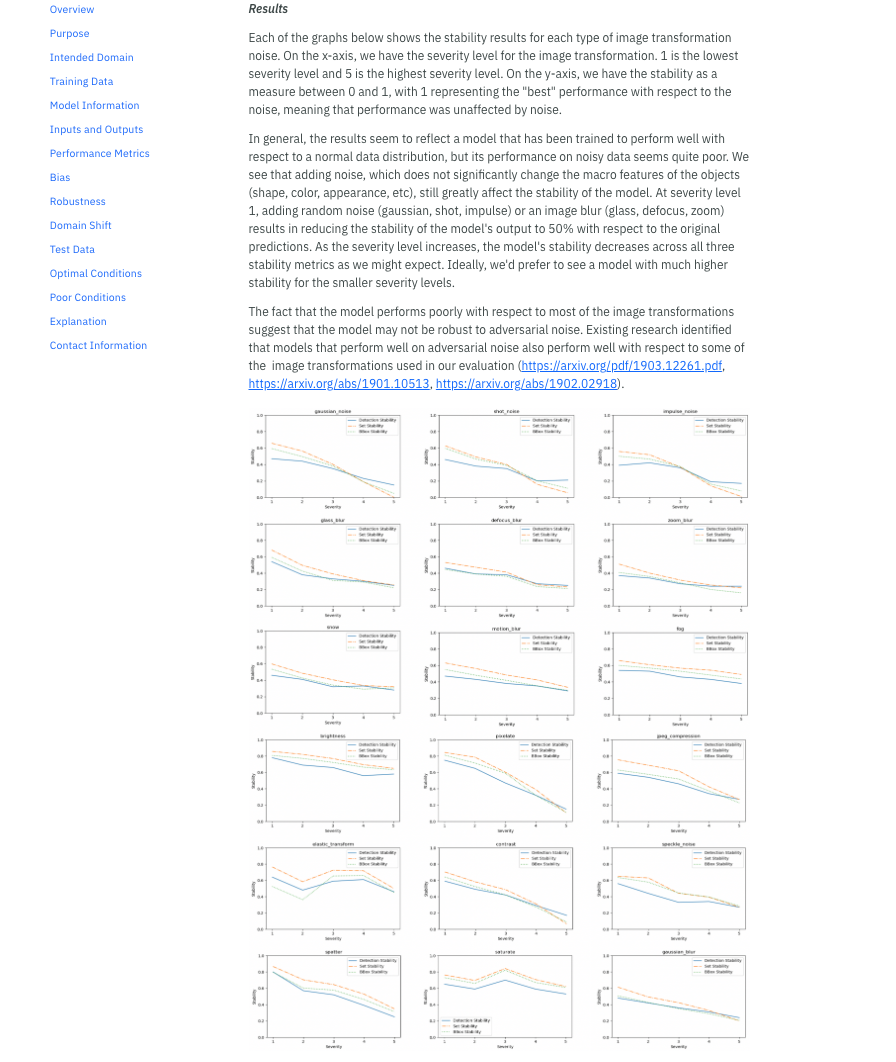}\hfil
\caption{Object Detector FactSheet - 6}
\end{figure*}

\newpage
\begin{figure*}
\centering
    \includegraphics[width=0.95\linewidth]{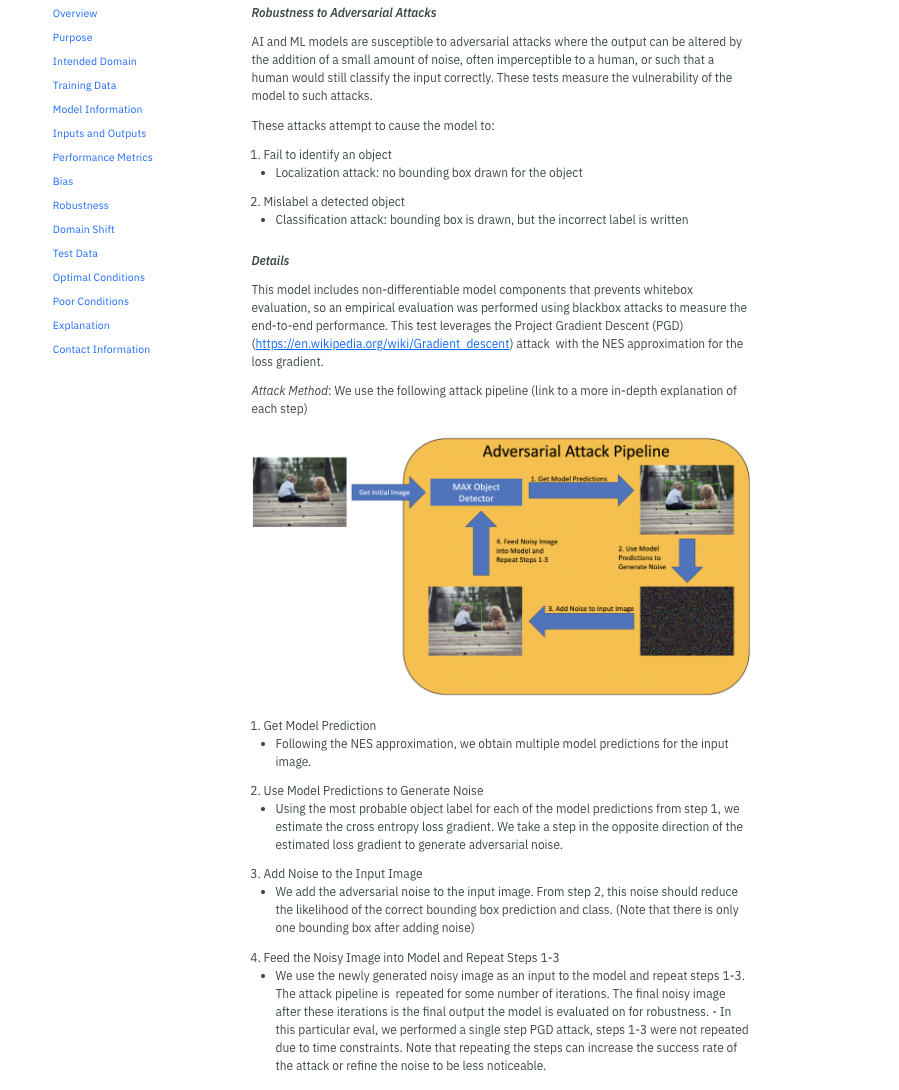}\hfil
\caption{Object Detector FactSheet - 7}
\end{figure*}

\newpage
\begin{figure*}
\centering
    \includegraphics[width=0.95\linewidth]{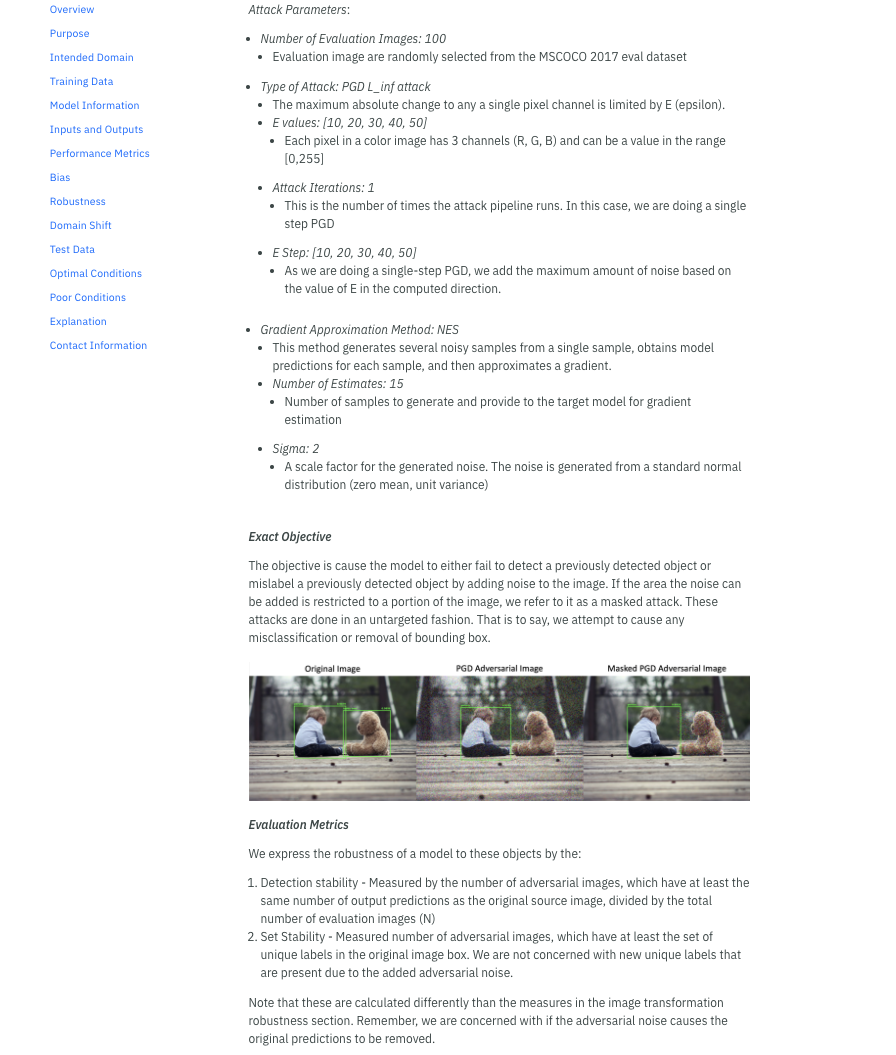}\hfil
\caption{Object Detector FactSheet - 8}
\end{figure*}

\newpage
\begin{figure*}
\centering
    \includegraphics[width=0.95\linewidth]{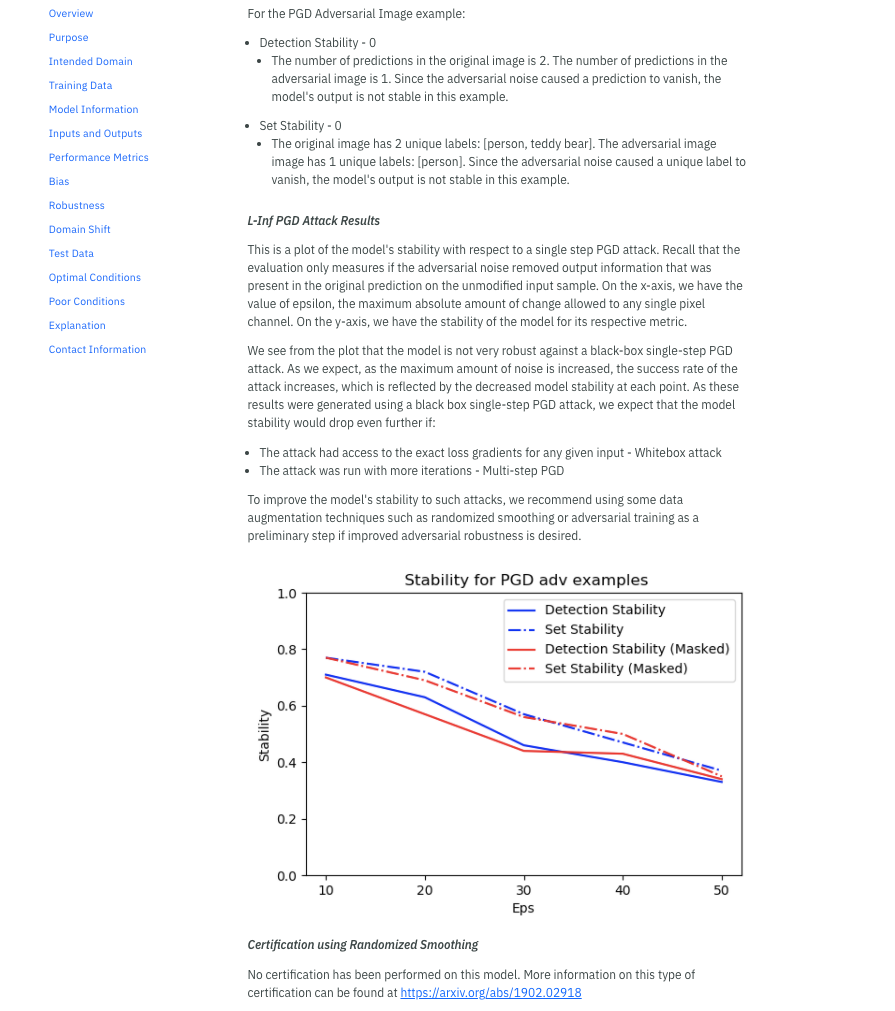}\hfil
\caption{Object Detector FactSheet - 9}
\end{figure*}

\newpage
\begin{figure*}
\centering
    \includegraphics[width=0.95\linewidth]{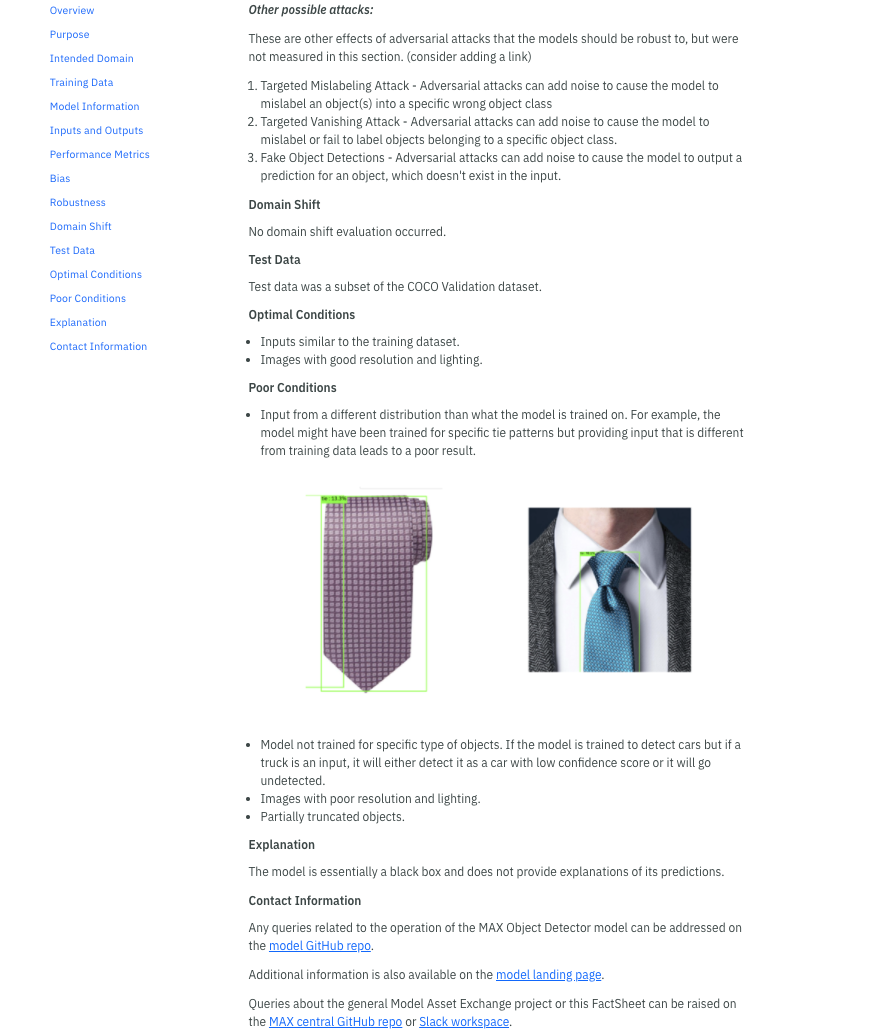}\hfil
\caption{Object Detector FactSheet - 10}
\end{figure*}

\newpage
\bibliographystyle{abbrv}
\bibliography{refs}

\end{document}